# Look-Ahead AC Optimal Power Flow: A Model-Informed Reinforcement Learning Approach

Xinyue Wang[1], Haiwang Zhong[1,2], Guanglun Zhang[1], Guangchun Ruan[1], Yiliu He[1], Zekuan Yu[1]
[1]Department of Electrical Engineering, Tsinghua University
[2]Sichuan Energy Internet Research Institute, Tsinghua University
Beijing 100084, China
zhonghw@mail.tsinghua.edu.cn

*Abstract*—With the increasing proportion of renewable energy in the generation side, it becomes more difficult to accurately predict the power generation and adapt to the large deviations between the optimal dispatch scheme and the day-ahead scheduling in the process of real-time dispatch. Therefore, it is necessary to conduct look-ahead dispatches to revise the operation plan according to the real-time status of the power grid and reliable ultra-short-term prediction. Application of traditional model-driven methods is often limited by the scale of the power system and cannot meet the computational time requirements of real-time dispatch. Data-driven methods can provide strong online decision-making support abilities when facing large-scale systems, while it is limited by the quantity and quality of the training dataset. This paper proposes a model-informed reinforcement learning approach for look-ahead AC optimal power flow. The reinforcement learning model is first formulated based on the domain knowledge of economic dispatch, and then the physics-informed neural network is constructed to enhance the reliability and efficiency. At last, the case study based on the SG 126-bus system validates the accuracy and efficiency of the proposed approach.

*Keywords—data-driven, model-driven, look-ahead economic dispatch, deep reinforcement learning, physics-informed neural network*

## I. INTRODUCTION

With the renewable energy resources gradually occupy the main proportion on the generation side, their variability poses great challenges to the operation and optimization of the system. A centralized optimization model is needed to solve the problem of reactive power caused by high penetration of renewable energy sources, which takes both active and reactive power optimization into account. Therefore, the look-ahead economic dispatch (ED) in this paper is based on the AC power flow model.

In the field of power system economic dispatch, the traditional model-driven methods usually analyze the physical model mechanism of components, construct mathematical models such as optimization objectives and network constraints, and then propose algorithms to solve them [1][2]. These methods take the advantages of rigorous theoretical derivation and clear mathematical model construction. However, this kind of method also requires many assumptions to simplify the model for improvement of the computational efficiency [3]. Currently, researches on the look-ahead economic dispatch prefer to utilize physical models to solve multi-period optimization problems only in the small-scale power systems.

This work was supported by the National Key Research and Development Program of China (NO.2020YFB0905904).

Regarding the dispatch of large-scale power systems, the model-driven methods cannot meet the time requirements of online decision-making due to the computational complexity. Recently, data-driven methods represented by artificial intelligence technology have shown the application potential in the field of power system economic dispatch [4]-[7]. Data-driven methods apply a generalized approach to mine the intrinsic information characteristics of the operation data, and construct a universal statistical model to learn and describe these characteristics, which can fit and generalize the mapping relationship between input and output in the dispatching process. On the decision-making stage, through using this kind of method, the dependence on the mathematical model construction of power system is reduced and the real-time performance of the look-ahead dispatch is guaranteed. Considering the time-varying characteristics of renewable energy and load, a dynamic economic dispatch method for integrated energy system was proposed in [8]. The authors in [9] introduced the deep deterministic policy gradient (DDPG) algorithm and utilized it to reduce the uncertainty of complex system configurations. Reference [10] formulated a dynamic economic dispatch model and combined imitation learning into it to fully make use of the uncertainty information in the operation data sources. Deep learning and reinforcement learning (RL) methods was employed in [11][12]. It can better adapt to the uncertain factors caused by renewable energy and fluctuating loads.

To date, most of the research that used data-driven methods mainly focused on the DC power flow model in look-ahead economic dispatch. The dimension of state quantity in AC power flow model is larger than that in DC model, which makes it more difficult and less efficient for neural network to extract effective information from the input state. Therefore, this paper proposes a model-informed reinforcement learning approach and construct the physics-informed neural network to refine the input information of the data-driven methods. The rest of this paper is organized as follows. Section II constructs the look-ahead economic dispatch model. Section III proposes a hybrid neural network training mode combining model-driven and data-driven methods. Case studies and conclusion are presented in Sections IV and V, respectively.

## II. MODEL FORMULATION OF THE LOOK-AHEAD ECONOMIC DISPATCH

The look-ahead economic dispatch problem is defined as a multi-period optimal generator output decision-making problem within a look-ahead time window considering security constraints [13][14]. In this section, the mathematical formulations of look-ahead economic dispatch model are established based on the AC power flow environment.

## A. Objective function

To simplify the formulation of the model, we indicate the time step by adding $\tau$ to all variables where $\tau$ varies from 1 to $T$, $T$ refers to the total number of time steps in a look-ahead time window. Meanwhile, variable t is introduced to express the sequence number of the look-ahead time window in the dispatch result. Using this convention, the objective function aiming at minimizing the total cost among all generators and all time steps within a look-ahead time window was given by:

$$\min \sum_{\tau=1}^{T} C(P_{\tau}^{G,t}) = \sum_{\tau=1}^{T} \sum_{i=1}^{N_{Gen}} C_i(P_{i,\tau}^{G,t}) \quad (1)$$

where $C_i$ is the cost of the $i^{th}$ generator, $P_{i,\tau}^{G,t}$ is the active power output of the $i^{th}$ generator at the $\tau^{th}$ time step within the look-ahead time window $t$. Furthermore, the number of buses, generators, and branches of the power system are defined as $N_{Bus}$, $N_{Gen}$, and $N_{Line}$, respectively.

## B. Constraints

This paper mainly studies the basic constraints and common constraints of power system in AC power flow environment, including generator constraints, voltage constraints and power flow constraints.

*1) Generator output limit*

The active power injection of generators cannot be greater than the upper limit or less than the lower limit.

$$P_i^{G\min} \leq P_{i,\tau}^{G,t} \leq P_i^{G\max}, \forall i, \tau \quad (2)$$

where $P_i^{G\min}$ and $P_i^{G\max}$ correspond to the lower and upper bounds of the active power output of the $i^{th}$ generator.

Moreover, when adjusting the voltage, the reactive power output of the generator is also prohibited from exceeding the limit.

$$Q_i^{G\min} \leq Q_{i,\tau}^{G,t} \leq Q_i^{G\max}, \forall i, \tau \quad (3)$$

where $Q_i^{G\min}$ and $Q_i^{G\max}$ correspond to the lower and upper bounds of the reactive power output of the $i^{th}$ generator. $Q_{i,\tau}^{G,t}$ is the reactive power output of the $i^{th}$ generator at the $\tau^{th}$ time step within the look-ahead time window $t$.

*2) Ramping limit*

For the first look-ahead time window ($t=1$), there is no need to consider the ramping limit between its first time step and the corresponding time step in the previous window. After the first look-ahead time window ($t>1$), it is necessary to consider the ramping limit between its first time step and the corresponding time step in the previous window $t-1$. And the active generator output at other time steps in the window also needs to meet the ramping constraints according to the time series.

$$\begin{cases} -R_i^{\min} \Delta t \leq P_{i,\tau}^{G,t} - P_{i,\tau}^{G,t-1} \leq R_i^{\max} \Delta t, \forall i, t>1, \tau=1 \\ -R_i^{\min} \Delta t \leq P_{i,\tau+1}^{G,t} - P_{i,\tau}^{G,t} \leq R_i^{\max} \Delta t, \forall i, \tau=1,...,T-1 \end{cases} \quad (4)$$

where $R_i^{\min}$ and $R_i^{\max}$ denote the ramping rate limit of the $i^{th}$ generator over time interval $\Delta t$.

*3) Power flow limit*

In this paper, the branch load rate $p$ calculated as the ratio of the current value to thermal limit is the criterion to judge whether the power flow exceeds the limit:

$$p_{j,\tau}^t = \frac{I_{j,\tau}^t}{I_j^{\max} + \varepsilon} < 1, \forall j, \tau \quad (5)$$

where $I_{j,\tau}^t$ is current of the $j^{th}$ branch at the $\tau^{th}$ time step within the look-ahead time window $t$. $I_j^{\max}$ is the thermal limit of the $j^{th}$ branch, $\varepsilon$ is a small constant value that avoids the denominator happens to be zero.

*4) Voltage limit*

The bus voltage in the system cannot be greater than the upper limit or less than the lower limit.

$$v_k^{\min} \leq v_{k,\tau}^t \leq v_k^{\max}, \forall k, \tau \quad (6)$$

where $v_k^{\min}$ and $v_k^{\max}$ correspond to the lower and upper bounds of the voltage of the $k^{th}$ bus, $v_{k,\tau}^t$ is the voltage of the $k^{th}$ bus at the $\tau^{th}$ time step within the look-ahead time window $t$.

## III. MODEL-INFORMED DEEP REINFORCEMENT LEARNING APPROACH

This section converts the mathematical formulations into the look-ahead economic dispatch model that allows the implementation of deep reinforcement learning. On this basis, a physics-informed neural network training framework is then constructed.

### A. Deep Reinforcement Learning Model

Reinforcement learning instructs the agent to learn how to act in an environment to obtain the maximum total reward. In the look-ahead ED problem, actions decided by the agent generally corresponds to the multi-period generator output in the look-ahead time window, which is a continuous variable decision-making problem. Therefore, simple reinforcement learning algorithms cannot meet the requirement that the output of the agent need to be continuous. In this paper, DDPG, a strategy learning method for continuous behavior, is adopted. DDPG trained by the deep learning method is improved on the basis of DQN algorithm [15], using the convolutional neural network to simulate the strategy function and Q function. Like other reinforcement learning algorithms, it is mainly composed of the agent, environment, state, action and reward.

*1) Agent and Environment*

The agent will decide the initial action corresponding to the generator output of all time steps within a look-ahead time window. The environment refers to the operating environment of the power system, which is mainly devoted to calculating AC power flow and giving feedback of rewards and the state of the next step to the agent.

*2) State space*

The state space represents the environmental information perceived by the agent and the changes brought on by the action. It is the basis for the agent to make decisions and evaluate the long-term benefits. Therefore, the design of the state space directly determines the convergence speed and final performance of the DDPG algorithm. In this section, state space is defined as a vector with ($N_{Bus}*T + 2N_{Gen}*T$) dimensions, including the normalized nodal load at all time steps in the look-ahead time window and the corresponding actions in the previous window:

$$\mathbf{s_t} = [\mathbf{L}_1^t, \mathbf{L}_2^t, ..., \mathbf{L}_T^t, \mathbf{a_{t-1}}] \quad (7)$$

$$\mathbf{L}_\tau^t = [\overline{L}_{1,\tau}^t, \overline{L}_{2,\tau}^t, ..., \overline{L}_{N_{Bus},\tau}^t] \quad (8)$$

where $\mathbf{L}_\tau^t$ is the vector of normalized nodal loads at the $\tau^{th}$ time step in the look-ahead time window $t$. The dimension of $\mathbf{L}_\tau^t$ is equal to $N_{Bus}$. $\overline{L}_{k,\tau}^t$ in (8) is the normalized system nodal load, which is computed as follows:

$$\overline{L}_{k,\tau}^t = \frac{2 \times L_{k,\tau}^t - (L_k^{\max} + L_k^{\min})}{L_k^{\max} - L_k^{\min}} \quad (9)$$

where $L_{k,\tau}^t$ is the nodal load of the $k^{th}$ bus at the $\tau^{th}$ time step in the look-ahead time window $t$. $L_k^{\max}$ and $L_k^{\min}$ correspond to maximum and minimum nodal load of bus $k$ in the whole load dataset.

*3) Action space*

In this section, the action space is defined as a vector with ($2N_{Gen}*T$) dimensions, including the initial normalized generator power output and adjustment of voltage at all time steps in the look-ahead time window. Without loss of generality, the reactive power of generators in this paper is regulated by voltage:

$$\mathbf{a_t} = [\mathbf{a}_1^t, \mathbf{a}_2^t, ..., \mathbf{a}_T^t, \Delta\mathbf{v}_1^t, \Delta\mathbf{v}_2^t, ..., \Delta\mathbf{v}_T^t] \quad (10)$$

$$\mathbf{a}_\tau^t = [\overline{a}_{1,\tau}^t, \overline{a}_{2,\tau}^t, ..., \overline{a}_{N_{Gen},\tau}^t] \quad (11)$$

$$\Delta\mathbf{v}_\tau^t = [\overline{\Delta v}_{1,\tau}^t, \overline{\Delta v}_{2,\tau}^t, ..., \overline{\Delta v}_{N_{Gen},\tau}^t] \quad (12)$$

where $\mathbf{a}_\tau^t$ is the vector of normalized active power outputs of the agent at the $\tau^{th}$ time step in the look-ahead time window $t$. $\Delta\mathbf{v}_\tau^t$ is the vector of voltage adjustment. Their dimensions both equal to the number of system generators $N_{Gen}$. $\overline{\Delta v}_{i,\tau}^t$ is the value of voltage adjustment corresponding to the $i^{th}$ generator at the $\tau^{th}$ time step in the look-ahead time window $t$. $\overline{a}_{i,\tau}^t$ in (11) is the normalized value of the active power output, which is computed as follows:

$$\overline{a}_{i,\tau}^t = \frac{2 \times a_{i,\tau}^t - (P_i^{G\max} + P_i^{G\min})}{P_i^{G\max} - P_i^{G\min}} \quad (13)$$

where $\overline{a}_{i,\tau}^t$ is the output of agent corresponding to the $i^{th}$ generator at the $\tau^{th}$ time step in the look-ahead time window $t$. Equation (13) lays the foundation for calculating the actual generator output in Section III-C.

*4) Reward*

The design of reward function can make the algorithm get the direction of policy update and finally converge to the optimal result. The reward is designed to be the objective function of the look-ahead ED model, added with penalties corresponding to the constraints in Section II-B:

$$r_t = -\sum_{\tau=1}^{T}\sum_{i=1}^{N_{Gen}} C_i(P_{i,\tau}^{G,t}) + \sum_{i=1}^{5} r_{pen}^i \quad (14)$$

To be consistent with the goal of maximizing the reward value, the objective of the look-ahead ED model is added by taking the inverse number. Moreover, five penalties are added in the reward to avoid possible constraint violations and instruct the agent to update in the direction of satisfying the security constraints.

- Penalty items for generator output and ramping limit:

$$r_{pen}^1 = w_1 \sum_{\tau=1}^{T}\sum_{i=1}^{N_{Gen}} (\max\{P_{i,\tau}^{G,t} - P_i^{G\max}, 0\} + \max\{P_i^{G\min} - P_{i,\tau}^{G,t}, 0\}) \quad (15)$$

$$r_{pen}^2 = w_2 \sum_{\tau=1}^{T}\sum_{i=1}^{N_{Gen}} (\max\{Q_{i,\tau}^{G,t} - Q_i^{G\max}, 0\} + \max\{Q_i^{G\min} - Q_{i,\tau}^{G,t}, 0\}) \quad (16)$$

$$r_{pen}^3 = w_3 \sum_{\tau=1}^{T}\sum_{i=1}^{N_{Gen}} (\max\{\Delta P_{i,\tau}^{G,t} - R_i^{\max}\Delta t, 0\} + \max\{-R_i^{\min}\Delta t - \Delta P_{i,\tau}^{G,t}, 0\}) \quad (17)$$

$$\Delta P_{i,\tau}^{G,t} = \begin{cases} 0 & ,\tau=1, t=1 \\ P_{i,\tau}^{G,t} - P_{i,\tau}^{G,t-1} & ,\tau=1, t\neq 1 \\ P_{i,\tau+1}^{G,t} - P_{i,\tau}^{G,t} & ,\tau\neq 1, t\neq 1 \end{cases} \quad (18)$$

- Penalty item for power flow limit:

$$r_{pen}^4 = w_4 \sum_{\tau=1}^{T}\sum_{j=1}^{N_{Line}} \max\{\frac{I_{j,\tau}^t}{I_j^{\max}+\varepsilon} - 1, 0\} \quad (19)$$

- Penalty item for voltage limit:

$$r_{pen}^5 = w_5 \sum_{\tau=1}^{T}\sum_{k=1}^{N_{Bus}} (\max\{v_{k,\tau}^t - v_k^{\max}, 0\} + \max\{v_k^{\min} - v_{k,\tau}^t, 0\}) \quad (20)$$

where $r_{pen}^i$ is the penalty for violating the $i^{th}$ constraint. $w_i$ corresponds to the weight factor of $r_{pen}^i$. In (18), $\Delta P_{i,\tau}^{G,t}$ is the variation of the active power output of the $i^{th}$ generator at the $\tau^{th}$ time step in the look-ahead time window $t$.

*B. Training of the physics-informed neural network*

At look-ahead time window $t$, the DDPG agent takes $\mathbf{a_t}$ and inputs it to the environment, transferring the environment state from $\mathbf{s_t}$ to $\mathbf{s_{t+1}}$, receiving an immediate reward $r_t$. Subsequently, the agent takes new action $\mathbf{a_{t+1}}$ according to certain strategies based on the environmental feedback $\mathbf{s_{t+1}}$. The training framework is shown in Fig. 1.

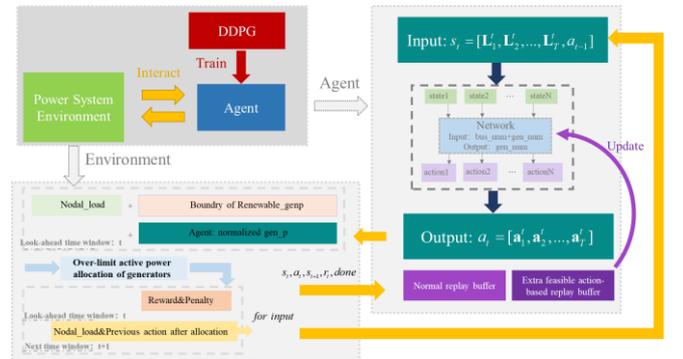

Fig. 1. Framework of training strategy for the look-ahead ED model based on DDPG

Here, the action output is normalized to ensure that the initial calculated active power after denormalization satisfies the constraint in Equation (2). However, the ramping limit constraints of thermal power generators cannot be guaranteed at the same time. Therefore, this paper designs the action security modification to fine-tune the generator active power, combining the adjustable capacity of each generator to carry out the proportional distribution of the deviation between total active power of generators and nodal loads based on the equal incremental rate principle. After the action security modification, the final active power output of generators will be given to the operating environment for AC power flow calculation. Moreover, the renewable generators are also considered in this paper. The upper and lower active power limits of the renewable generators in Equation (2) become:

$$\begin{cases} P_i^{G\min} = 0 \\ P_i^{G\max} = P_{i,\tau}^{New\max,t} \end{cases} \quad (21)$$

where $P_{i,\tau}^{New\max,t}$ is the real-time upper bound of the active power output of the $i^{th}$ renewable generator at the $\tau^{th}$ time step within the look-ahead time window $t$. It is provided by the system operating environment.

The economic action security modification mainly includes the following three steps:

1) De-normalization of the action and calculation of the adjustable capacity

The flowchart of the de-normalization and adjustable capacity calculation is shown in Fig. 2. $\overline{a}_{i,\tau}^{t}$ is the agent output of the $i^{th}$ generator at the $\tau^{th}$ time step in the look-ahead time window $t$ and $a_{i,\tau}^{t}$ is the denormalized value. $V_{i,\tau}^{u,t}$ and $V_{i,\tau}^{l,t}$ correspond to the upper and lower adjustable active power capacity of the $i^{th}$ generator at the $\tau^{th}$ time step in the look-ahead time window $t$.

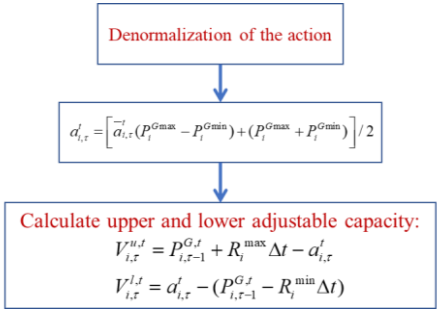

Fig. 2. Flowchart of the denormalization and adjustable capacity calculation

2) Redistribution of the over-limit active power output

The flowchart of the over-limit active power output redistribution is shown in Fig. 3. In this paper, ramping limit constraints is only incorporated for thermal generators. If $a_{i,\tau}^{t}$ of the thermal generator exceeds the upper bound, it will be binded to the upper boundary, and the upper adjustable capacity $V_{i,\tau}^{u,t}$ will be set zero accordingly. The case where $a_{i,\tau}^{t}$ exceeds the lower bound can also be handled similarly.

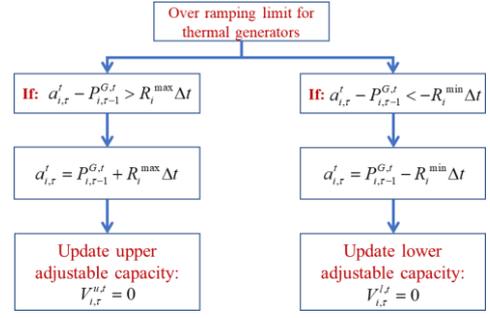

Fig. 3. Flowchart of the over-limit active power output redistribution

3) Elimination of the deviation based on the equal incremental rate principle

The flowchart of the deviation elimination based on the equal incremental rate principle is shown in Fig. 4. If the deviation value is greater than a set constant $\sigma$, the active power output $a_{i,\tau}^{t}$ will be distributed until it meets the requirements. $\Delta P_{\tau}^{sum,t}$ is the deviation between total active power of generators and nodal loads at the $\tau^{th}$ time step in the look-ahead time window $t$. The constant value of maximum iteration times is $M$. The constant value $\sigma$ equals to 1e-8. $k_{i,\tau}^{t}$ is the differential value of generation cost function at $a_{i,\tau}^{t}$. $P_{i,\tau-1}^{G,t}$ is the final generator active power which will be fed back to the system operating environment.

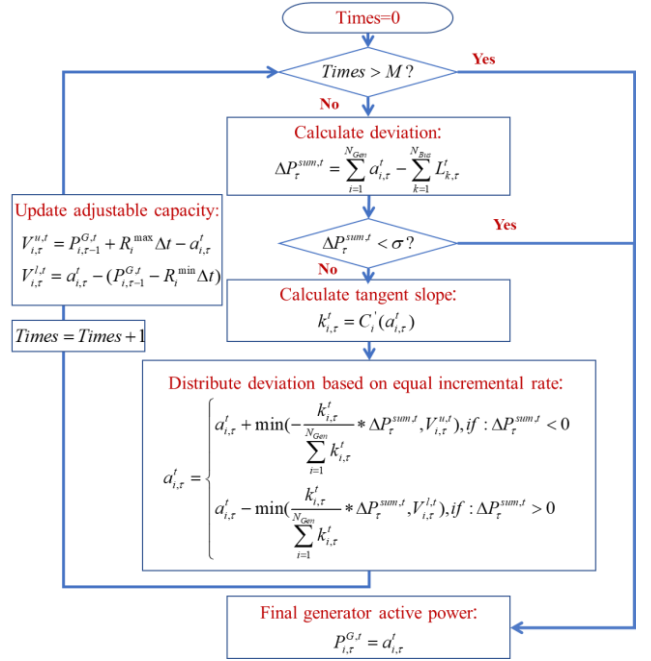

Fig. 4. Flowchart of the deviation elimination based on equal incremental rate

IV. CASE STUDY

A. Simulation Setup

The case study is conducted on a laptop with Intel i9 10900K. The programming environment is python 3.6 with pypower 5.1.15 and torch 1.9.0. The proposed model-informed reinforcement learning approach is tested on a

SG126-bus system, which considers AC power flow calculation. The number of buses, generators, and branches are shown in Table I.

TABLE I
POWER SYSTEM PARAMETERS

| System name | Number of buses | Number of generators | Number of branches |
|---|---|---|---|
| SG126 | 126 | 54 (18 renewable) | 185 |

The cost curve of the generators is illustrated in Fig. 5. Without loss of generality, the cost function of generators is assumed to be a quadratic function and there are 10 typical cost functions in the SG126-bus system.

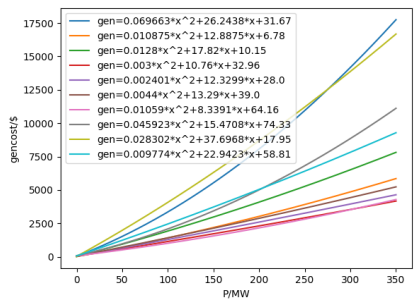

Fig. 5. Cost curve of generators

### B. Simulation results

Both the actor and critic networks are composed of multi-layer perceptron (MLP) with hidden layers. The actor network contains 5 layers and its learning rate is 1e-7. The critic network contains 3 layers with the learning rate of 1e-3. To make the model fit the realistic operation scenario of the power system, a hard overload setting is added to the security constraint of the branch power flow limit. If the loading rate of the branch in Equation (6) exceeds 1.4, the corresponding branch will be disconnected, which may cause the AC power flow calculation of the system diverge. Once the power flow calculation does not converge, the agent will stop exploration of next steps in the current episode. Meanwhile, for the episodes that can explore all steps, their reward will plus a positive constant to encourage the strategy to learn in these directions.

a) Training stage

Three training modes are compared:
M1: without action security modification;
M2: modification only following the adjustable capacity;
M3: the economic modification based on the equal incremental rate principle proposed in this paper.

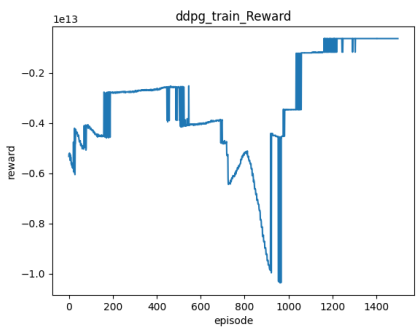

Fig. 6. Reward curve of the training process in M1

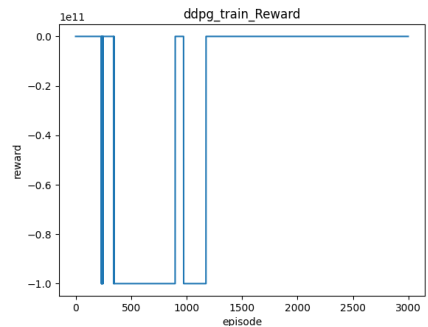

Fig. 7. Reward curve of the training process in M2

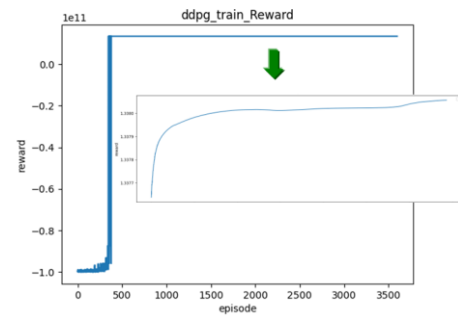

Fig. 8. Reward curve of the training process in M3

In the training process, it can be observed that without the over-limit action security modification, the active output of the generators determined by the agent easily cause the divergence of the AC power flow calculation, and the final reward corresponding to the convergence of the curve is a large negative value, which means the training results cannot meet the security constraints. The reward curve of the training process in M2 also shows continuous oscillation, and its final reward value is lower than M3. In contrast, based on equal incremental rate, the proposed M3 performs well in convergence and achieves a higher reward score without penalty items.

The training dataset of the look-ahead ED model is monthly load profile, and two-week load profile is used for online test. In this case, each look-ahead time window spans over 4 hours with a temporal resolution of 15 minutes. $\Delta t$ is 15 minutes and $T$ is 16. We use the bus-wise net load predictions at each time step in the look-ahead window and take the first time step's generator active power into actual operation. Corresponding to the testing load profile, the number of time windows is 1344 and the online decision time is 0.17s per window.

b) Testing stage

Under the same load profile, the training effects of the agents corresponding to the above three training modes are tested.

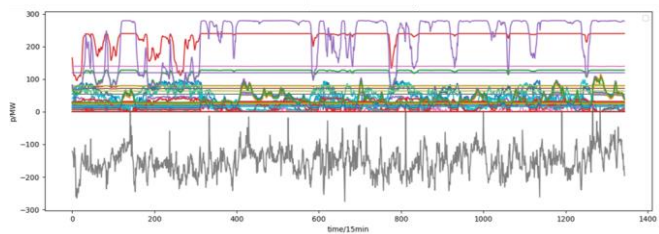

Fig. 9. The actual generator power output of the look-ahead ED in M1

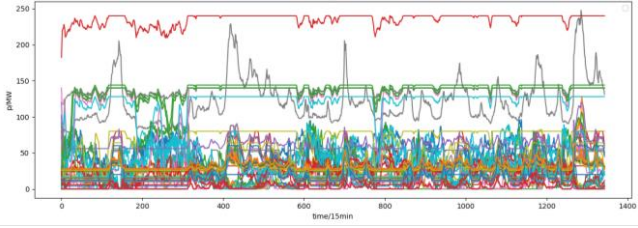

Fig. 10. The actual generator power output of the look-ahead ED in M2

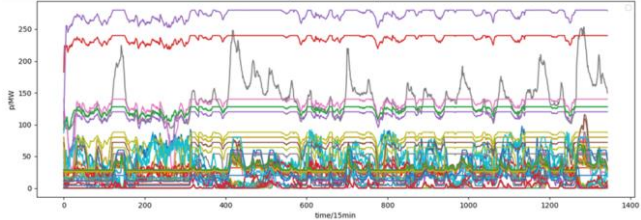

Fig. 11. The actual generator power output of the look-ahead ED in M3

The grey curve with negative value in Fig. 9 shows that the power output fails to meet the generator output limit constraint, which is not allowed in actual system operation. The convergent AC power flow calculation results in Fig. 10 and Fig. 11 both satisfy all the security constraints in Section II. To compare the economic efficiency between them, the traditional optimal power flow calculation results under the same load profile is taken as the benchmark.

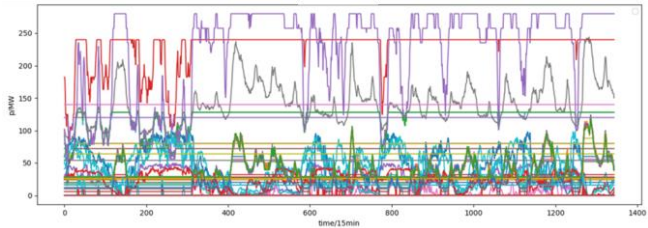

Fig. 12. The actual generator power output of traditional OPF

It can be found that the trend of the generator output curve obtained by M3 is similar to that of OPF. And the detailed efficiency comparison is shown in Table II.

TABLE II
COMPARISON OF TRAINING MODES

| Name | Security constraints | Training Time (h) | Economic efficiency (compared to OPF) |
|---|---|---|---|
| M1 | × | 10 | × |
| M2 | √ | 18 | +12.9% |
| M3 | √ | 20 | +4.1% |

The testing result of the proposed approach has an error of about 4.10% compared with the total cost of OPF, which is lower than M2. It is demonstrated that the proposed M3 better in accuracy.

## V. CONCLUSION

This paper proposes a model-informed reinforcement learning approach for look-ahead ED. A physics-informed neural network is then constructed based on the DDPG algorithm to ensure the convergence of the agent. In the training process, active power outputs is fine-tuned in the action security modification, where the adjustable capacity of each generator is calculated to carry out the proportional distribution of the deviation based on the equal incremental rate principle. Simulation results validate advantage of the proposed look-ahead ED model.

However, in the larger-scale system, due to the sharp increase in the number of buses, generators and branches, the training efficiency and the rationality and robustness of the result need further investigating.